\documentclass{jfm}
\usepackage{graphicx} 
\usepackage{graphicx}
\usepackage{epstopdf, epsfig}
\usepackage{color}

\newcommand{\be}{\begin{equation}}
\newcommand{\ee}{\end{equation}} 
\newcommand{\bea}{\begin{eqnarray}}
\newcommand{\eea}{\end{eqnarray}}

\title{Sweeping effect and Taylor's hypothesis  via correlation function}

\author{MAHENDRA K. VERMA\aff{1} and ABHISHEK KUMAR\aff{1} \corresp{\email{abhishek.kir@gmail.com}}}
  
\affiliation{\aff{1}Department of Physics, Indian Institute of Technology, Kanpur, India 208016}

\begin{document}

\maketitle

\begin{abstract}
We performed high-resolution numerical simulations of hydrodynamic turbulence with and without mean velocity ($U_0=0,10$), and demonstrate the sweeping effect.  For $U_0=0$, the velocity correlation function, $C({\bf k},\tau)$ decays with time due to eddy viscosity, but it also shows fluctuations due to the sweeping effect.  For $U_0=10$, $C({\bf k},\tau)$ exhibits damped oscillations with the frequency of $U_0 k$ and decay time scale corresponding to the $U_0=0$ case.  A closer examination of $\Im[C({\bf k},\tau)]$ also demonstrates sweeping effect for $U_0=10$.  We also demonstrate that the frequency spectra of the velocity fields measured by real-space probes are respectively $f^{-2}$ and $f^{-5/3}$ for $U_0=0$ and 10; these spectra are related to the Lagrangian and Eulerian space-time correlations. 
\end{abstract}

\section{Introduction}
The incompressible Navier--Stokes equation of a flow that is moving with a mean velocity of ${\mathbf U}_0$ is
\begin{eqnarray}
\partial_t u_i + ({\mathbf U}_0 \cdot \nabla) u_i+ \partial_j (u_j u_i)  & = &  -\partial_i p + \nu \partial^2 u_i + f_i, 
\label{eq:NS1} \\
\partial_i u_i  & = & 0,  \label{eq:NS2}
\end{eqnarray}
where ${\bf u}$ is the velocity fluctuation with a zero mean, ${\bf f}$ is the external force,  $p$ is the pressure, and $\nu$ is the kinematic viscosity.   One of the most important principles of classical physics is Galilean invariance, according to which laws of physics are the same in all inertial frames (frames moving with constant velocities with relative to each other). Naturally, the Navier--Stokes equation, which is Newton's laws for fluid flows, exhibits this symmetry~\citep{Lesieur:book:Turbulence,Frisch:book,Davidson:book:Turbulence,McComb:book:Turbulence,McComb:book:new}.   As a consequence of this symmetry, the flow properties of the fluid in the laboratory reference frame (in which the fluid moves with a mean velocity of ${\mathbf U}_0$) and in the co-moving reference frame  (${\mathbf U}_0=0$) are the same

The velocity field of a turbulent flow is random, hence it is typically characterised by its correlations. There have been several major advances in the understanding the correlations in homogeneous and isotropic turbulence, most notably by Kolmogorov~\citep{Kolmogorov:DANS1941Structure,Kolmogorov:DANS1941Dissipation} who showed that in the inertial range, the velocity correlation $C({\bf k}) = K_\mathrm{Ko} \Pi^{2/3} k^{-5/3}/(4\pi k^2)$, where $\Pi$ is the energy flux, and $K_\mathrm{Ko}$ is the Kolmogorov constant.  The corresponding one-dimensional energy spectrum is $ E(k) = K_\mathrm{Ko} \Pi^{2/3} k^{-5/3}$.   

\citet{Kraichnan:PF1964Eulerian} argued that in the presence of random ${\mathbf U}_0$, Eulerian field theory does not yield  Kolmogorov's spectrum.  In particular, \citet{Kraichnan:PF1964Eulerian} considered a  fluid flow with a {\em random} mean velocity field that is constant in space and time but has a Gaussian and isotropic distribution over an ensemble of realisations.  Then he employed direct interaction approximation (DIA) to close the hierarchy of equations and showed that $E(k) \sim (\Pi U_\mathrm{0})^{1/2} k^{-3/2}$, where $U_\mathrm{0}$ is the rms value of the mean velocity.    \citet{Kraichnan:PF1964Eulerian} argued that the above deviation of the energy spectrum from the experimentally observed Kolmogorov's $k^{-5/3}$ energy spectrum is due to the {\em sweeping effect} according to which  small-scale fluid structures are advected by the large energy-containing eddies.   Due to the above observations, Kraichnan  emphasised that the Eulerian formalism is inadequate for obtaining Kolmogorov's spectrum for a fully developed fluid turbulence. Later, he developed Lagrangian field theory of hydrodynamic turbulence that is consistent with the Kolmogorov's 5/3 theory of turbulence ~\citep[see][and other related papers]{Kraichnan:PF1965Lagrangian_history}. The above framework is called  {\em random Galilean invariance}. 

A related phenomenon is Taylor's hypothesis of frozen turbulence. \citet{Taylor:PRS1938} proposed that the velocity measurement at a point in a fully-developed turbulent flow moving with a constant velocity ${\bf U}_0$  (e.g. in a wind tunnel)  can be used to study the velocity correlations.  This is because the mean flow  advects the {\em frozen-in} fluctuations, and the stationary probe in the fluid measures the fluctuations along a line.  Here, the frequency spectrum of the measured time series is expected to show $\omega^{-5/3}$, where $\omega$ is the angular frequency.    This proposal, {\em Taylor's frozen-in turbulence hypothesis}, has been used in many experiments to ascertain Kolmogorov's spectrum. 

In this paper, we investigate the sweeping effect as well as Taylor's hypothesis using numerical simulations.   Using  numerical data, we compute the normalised  correlation function $R(\mathbf k, \tau)$, defined as 
\begin{equation}
R(\mathbf k, \tau) = \frac{C(\mathbf k, \tau)}{C(\mathbf k, 0)} = \frac{\langle \mathbf u(\mathbf k, t)\cdot \mathbf u^{*}(\mathbf k, t+\tau) \rangle}
{\langle |\mathbf u(\mathbf k, t)|^2 \rangle} ,
  \label{eq:R}
\end{equation} 
where $\mathbf  u(\mathbf k, t)$ is the Fourier transform of the velocity field $\mathbf u(\mathbf  x,t)$.  This measure was earlier proposed by ~\citet{Sanada:PF1992}. For $U_0=0$,  theoretical calculations~\citep{Yakhot:JSC1986,McComb:book:Turbulence} predict that 
\be
R(\mathbf k, \tau) = \exp(-\tau/\tau_c(k)),
\ee
where $\tau_c$ is decay time of an eddy of size $1/k$.  The decay time  based on local velocity is
\be
\tau_c(k) \approx \frac{1}{k u_k} \sim \Pi^{-1/3} k^{-2/3},
\ee
but the decay time based on sweeping by random mean velocity~\citep{Kraichnan:PF1964Eulerian} is
\be
\tau_c(k) \approx \frac{1}{k U_0} \sim (U_0 k)^{-1}.
\ee
\citet{Sanada:PF1992} performed numerical spectral simulations on $256^3$ grid resolution and computed $\tau_c(k)$.  They observed that $\tau_c$ is closer to $k^{-1}$ than $k^{-2/3}$, thus arguing in favour of sweeping effect  proposed by  \citet{Kraichnan:PF1964Eulerian}.

In this paper, we compute $R(\mathbf k, \tau) $ using numerical simulations and find this to be a  complex number whose phase evolution can be related to the advection by random velocity field. We model
\be
R(\mathbf k, \tau) = \exp(-\tau/\tau_c(k)) \exp(i \tilde{\bf U}_0 \cdot {\bf k} \tau),
 \label{eq:Rk_tau}
\ee
where $\tilde{\bf U}_0$ is the random mean velocity.   Thus we argue that the effect of the random mean velocity appears in the phase of $R(\mathbf k, \tau) $, rather than on the absolute value of $R(\mathbf k, \tau) $ as argued by \citet{Sanada:PF1992}.  Using this approach, we validate Kraichnan's random Galilean invariance.
 
For finite $U_0$ and $U_0 \gg \tilde{U}_0$, we compute $R(\mathbf k, \tau) $ and observe that 
\be
R(\mathbf k, \tau) \sim \exp(-\tau/\tau_c) \exp(i {\bf U}_0 \cdot {\bf k} \tau + i \tilde{\bf U}_0 \cdot {\bf k} \tau).
 \label{eq:Rk_tau_withU0}
\ee
Thus, an eddy is advected by the mean flow as well as by random mean velocity.  For this case, the frequency spectrum of the velocity probe in real space  varies as $\omega^{-5/3}$, thus verifying Taylor's frozen-in turbulence hypothesis~\citep{Taylor:PRS1938}.  When $U_0=0$, the corresponding frequency spectrum is $\omega^{-2}$.  We contrast these two cases using correlation function.

In the next section, we briefly describe the sweeping effect and demonstrate its signature using numerical simulation. In \S\ref{sec:f-2}, we show that the frequency spectrum $E(f)\sim f^{-2}$ for turbulent flow in the absence of a constant mean velocity field $U_0$, where $f=\omega/2\pi$. In \S\ref{sec:Taylor}, we discuss Taylor's frozen-in turbulence hypothesis and show $E(f)\sim f^{-5/3}$ for $U_0 \gg \tilde{U}_0$. In \S\ref{sec:elliptic}, we revisit the elliptic approximation~\citep{He:PRE2010,He:PRE2011,He:ARFM2016} in terms of correlation function discussed in \S\ref{sec:Taylor}. We conclude in \S\ref{sec:conclusion}.

\section{Sweeping effect and its numerical verification}
\label{sec:sweeping_effect}

Here we briefly describe the {\em sweeping effect}, first proposed by \citet{Kraichnan:PF1964Eulerian}. 
Kraichnan assumed that the velocity fluctuation of Navier--Stokes equation is advected by the mean flow, ${\bf U}_0$, and  {\em random large-scale flow}, $\tilde{\bf U}_0$.  Hence, the temporal evolution of  fluctuating Fourier mode ${\bf u}({\bf k})$ is given by
\be
\frac{\partial {\bf u}({\bf k})}{\partial t} = -i [{\bf k} \cdot ({\bf U}_0 + \tilde{\bf U}_0)] {\bf u}({\bf k}).
\ee
 Kraichnan assumed that $\tilde{\bf U}_0$ is spatially varying but is constant in time.  Under the assumption of Gaussian distribution for $\tilde{\bf U}_0$, \citet{Kraichnan:PF1964Eulerian} \citep[also see][]{Wilczek:PRE2012} showed that  
 \bea
R({\bf k}, \tau) & = &  \exp[-i {\bf k} \cdot {\bf U}_0 \tau] \langle \exp[-i {\bf k} \cdot \tilde{\bf U}_0 \tau]  \rangle \\
 & = &  \exp[-i {\bf k} \cdot {\bf U}_0 \tau - \frac{\langle \tilde{U}_0^2 \rangle k^2 \tau^2}{6}].  
 \label{eq:R_Kraichnan}
 \eea
When we compare the above equation with Eq.~(\ref{eq:Rk_tau}), we observe that Kraichnan's derivation does take into account the damping factor $\exp(-\tau/\tau_c)$.

 In the following discussion, using numerical data, we will compute $R({\bf k}, \tau)$  that provides signatures of the sweeping effect.   Since our analysis is based on the Eulerian framework, it is necessary to review the relevant results of Eulerian field theory that has been employed to analyse the turbulent velocity field.  \citet{Kraichnan:JFM1959} employed direct interaction approximation (DIA), while \citet{Yakhot:JSC1986}, McComb and coworkers~\citep{McComb:book:Turbulence}, \citet{DeDominicis:PRA1979}, \citet{Zhou:PR2010} employed renormalisation group analysis in the Eulerian field-theoretic framework.  This is an exhaustive field, and it has been reviewed by  \citet{McComb:book:Turbulence,McComb:book:new} and \citet{Zhou:PR2010}.  Also, see Appendix~\ref{sec:rgcalc}. For the following discussion, it suffices to remark that the  {\em dressed Green's function} for $U_0=0$ in the turbulent regime is
\be
G({\bf k},\omega) = \frac{1}{-i\omega + \nu(k) k^2},
\label{eq:Gk}
\ee
where
\be
\nu(k) = \nu_* \sqrt{K_\mathrm{Ko}} \epsilon^{1/3} k^{-4/3}
\label{eq:nu_k}
\ee
 is the turbulent viscosity.  Here $\nu_*$ is a constant whose value is approximately $0.38$~\citep{McComb:book:Turbulence} (also see Appendix~\ref{sec:rgcalc}).   Transformation of the above function to the temporal space yields
\be
G({\bf k},\tau) =\theta(\tau) \exp{[- \nu(k) k^2 \tau)]},
\label{eq:Gkt_real}
\ee
where $\theta(\tau)$ is the step function.   The normalised correlation function $R({\bf k},\tau)$ (defined in Eq.~(\ref{eq:R})) too is time dependent, and it is assumed to have the same relaxation time as the Green's function, i.e.,
\be
R({\bf k},\tau) =\theta(\tau)  \exp{[- \nu(k) k^2 \tau]}.
\label{eq:Rkt_real}
\ee 
The above equations indicate that relaxation time scale for the correlation and Green's functions are
\be
\tau_c = \frac{1}{\nu(k) k^2} \sim \frac{1}{\epsilon^{1/3} k^{2/3}},
\label{eq:tauc_Euler}
\ee 
where $\nu(k)$ is given by Eq.~(\ref{eq:nu_k}).  This is a generalisation of fluctuation dissipation theorem~\citep{McComb:book:Turbulence}. Using the numerically computed $C({\bf k},\tau)$, we estimate the relaxation time and compare it with the predictions of Eulerian field theory.

We perform numerical simulation of  Navier--Stokes equation in the turbulent regime for the mean velocity $\mathbf U_0=0$.  We employ pseudospectral code Tarang~\citep{Verma:Pramana2013tarang}  to simulate the flow  on a $512^3$ grid with random forcing.  We use the fourth-order Runge Kutta (RK4) scheme for  time stepping, 2/3 rule for dealiasing, and CFL condition for computing $\Delta t$.  The Reynolds number of the runs  are $u_\mathrm{rms} L/\nu =5.7 \times 10^3$, where $u_\mathrm{rms}$ is the rms value of the velocity fluctuations.  

\begin{figure}
\begin{center}
\includegraphics[scale = 1]{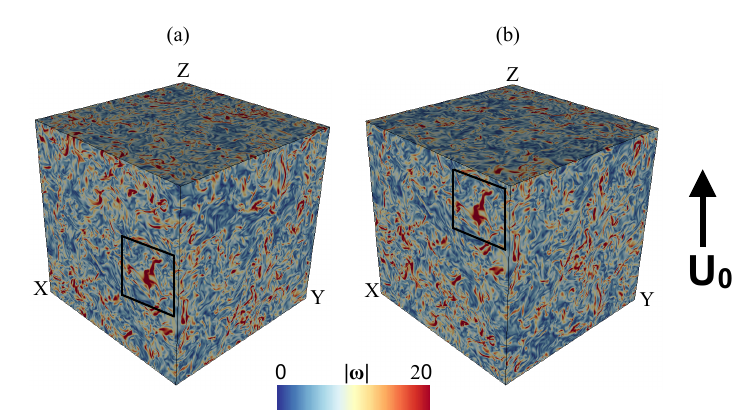}
\end{center}
\caption{A density plot of the magnitude of the vorticity $|\boldsymbol{\omega}|$ at $t = 0.2$ for (a) $\mathbf U_0 = 0$ and (b) $\mathbf U_0 = 10\hat{z}$.  The flow structures in the boxed zone of (b) are shifted by $\Delta z =U_0 t =  10\times 0.2 = 2$ units compared to (a), thus verifying Galilean invariance of the fluid equation. }
\label{fig:real_space}
\end{figure}

\begin{figure}
\begin{center}
\includegraphics[scale = 0.7]{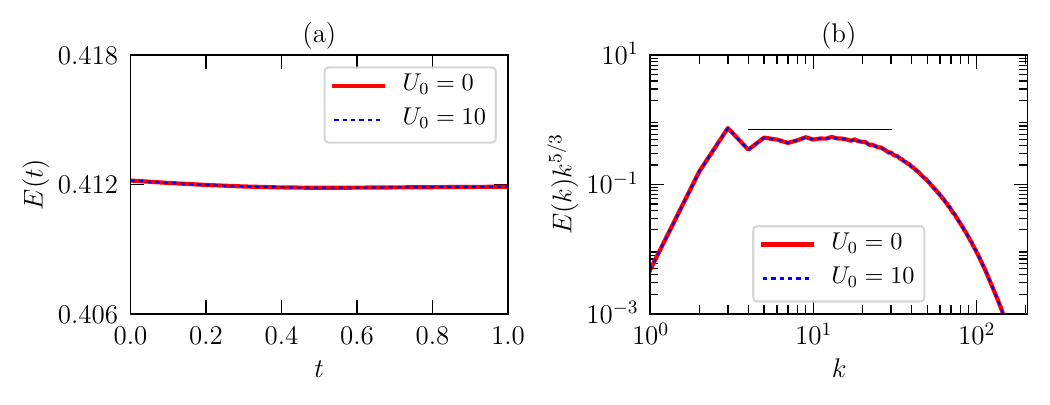}
\end{center}
\caption{For $U_0=0$ and $U_0=10$, the plots of (a) total energy of the velocity fluctuation, $u^2/2$, vs.~$t$ and (b) the normalized kinetic energy spectrum $E(k)k^{5/3}$ vs.~$k$.    The plots show that $E(t)$ and $E(k)$ are identical for $U_0=0,10$, thus verifying Galilean invariance of the fluid equation.}
\label{fig:energy_spectrum}
\end{figure}

We evolve the flow with $U_0=0$ till a steady state is reached.  At this point, we fork two simulations with ${\mathbf U}_0 = 0$ and ${\mathbf U}_0=10 \hat{z}$, and run it for one eddy turnover time.  In Fig.~\ref{fig:real_space}(a,b) we illustrate velocity profiles of the flows for $U_0=0$ and 10 at $t=0.2$.    The flow profiles  are identical except that the flow for ${\mathbf U}_0=10 \hat{z}$ is shifted vertically by $10 \times 0.2 = 2$ units, as expected.  For  $U_0=0$ and 10,  the  temporal evolution of the fluctuating energy, as well as the energy spectra, are identical, as illustrated in Fig.~\ref{fig:energy_spectrum}.

These observations on the flow evolution and energy spectra are on expected lines and they illustrate that the Navier--Stokes equation is Galilean invariant.  These results, however, are based on equal-time correlations; the subtleties however surface when we study the temporal correlation of the velocity modes.

\begin{figure}
\begin{center}
\includegraphics[scale = 0.7]{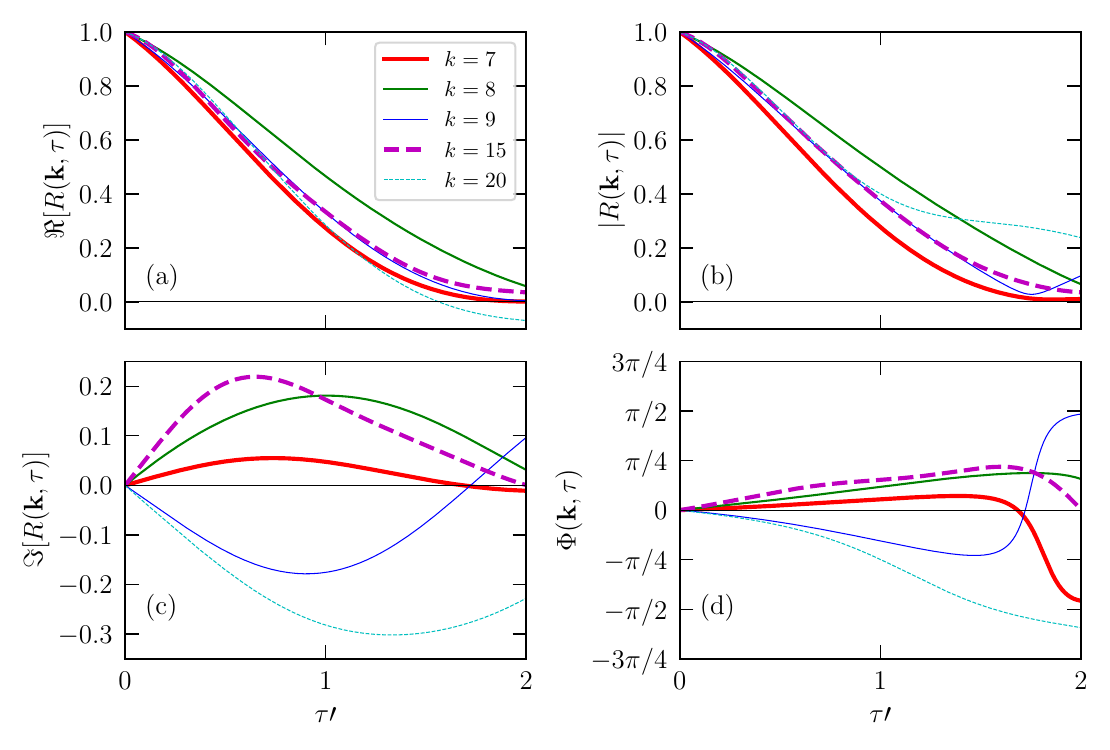}
\end{center}
\caption{For $\mathbf U_0 = 0$ and  $k = 7,8,9,15,20$ (inertial range wavenumbers), plots of the normalised correlation function $R(\mathbf k, \tau)$ vs.~$\tau'=\tau/\tau_c$:  (a) $\Re[R(\mathbf k, \tau)]$, (b) $|R(\mathbf k, \tau)|$, (c) $\Im[R(\mathbf k, \tau)]$, and (d) $\Phi(\mathbf k, \tau)$.  The real part and the absolute value decay exponentially in time  as Eq.~(\ref{eq:Rk_tau}), while the oscillating imaginary part and monotonic increase of  the $\Phi(\mathbf k, \tau)$ with time demonstrate the sweeping effect.}
\label{fig:corr}
\end{figure}

\begin{figure}
\begin{center}
\includegraphics[scale = 0.75]{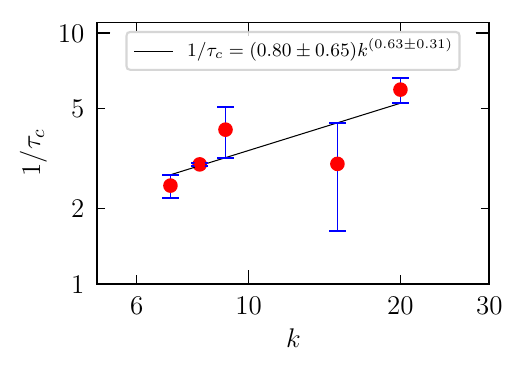}
\end{center}
\caption{Plot  of $\tau_c^{-1} $ vs.~$k$.   We observe that  $\tau_c^{-1} \sim  k^{0.63 \pm 0.31}$. The exponent being closer to 2/3 indicates that  Eq.~(\ref{eq:tauc_Euler}) is a fair description of the decaying time scale.}
\label{fig:tau_c_k}
\end{figure}

Using the numerical data, we compute the normalised  correlation function $R(\mathbf k, \tau)$ of Eq.~(\ref{eq:R}), and in Fig.~\ref{fig:corr} we plot  its real and imaginary parts, $\Re[R(\mathbf k, \tau)]$ and $\Im[R(\mathbf k, \tau)]$, as well as its magnitude, $|R(\mathbf k, \tau)|$, and phase 
\begin{equation}
\Phi(k,\tau) = \tan^{-1} \frac{\Im[R(\mathbf k, \tau)]}{\Re[R(\mathbf k, \tau)]s}
\end{equation}
for  $k = 7,8,9,15$ and 20, which lie in the inertial range.  The plot of Fig.~\ref{fig:corr}(a,b) show that  $\Re[R(\mathbf k, \tau)]$ and $|R(\mathbf k, \tau)|$  decay  exponentially with time, consistent with Eq.~(\ref{eq:Rk_tau}).  In Fig.~\ref{fig:tau_c_k}, we plot $\tau_c^{-1}$ vs.~$k$, where $\tau_c$ is obtained from the slope of semi-logy plot $|R(\mathbf k, \tau)|$ vs.~$\tau$.  A regression analysis yield  $\tau_c  \sim k^{-0.63\pm 0.31}$, which has significant error bar due to a limited range of data.  This is due to the relatively narrow inertial range of our $512^3$ grid simulations.  However, the slope of 0.63 is closer to the exponent of 2/3  than 1, which is contrary to the results of \citet{Sanada:PF1992}.  This result indicates that the decay of $R(\mathbf k, \tau)$  is described by Eq.~(\ref{eq:Rkt_real}).  In the following discussion, we show that $R(\mathbf k, \tau)$ is a complex number, and its phase contains information about the sweeping effect.
 
 In Figs.~\ref{fig:corr}(c,d)  we plot $\Im[R(\mathbf k, \tau)]$ and the phase $\Phi({\mathbf k},\tau)$ of $R(\mathbf k, \tau)$.  The nonzero values of $\Im[R(\mathbf k, \tau)]$ and the phase $\Phi({\mathbf k},\tau)$ indicates that we need to add a new component to Eq.~(\ref{eq:Rkt_real}).  The phases for various $k$'s have the following properties:
\begin{enumerate}
\item  The phase increases linearly with time till $\tau \approx \tau_c$, hence $\Phi({\mathbf k},\tau) \propto \tau$ till $\tau \approx \tau_c$.
\item  In Fig.~\ref{fig:corr}(d), the slopes of the $\Phi({\mathbf k},\tau)$ for various $k$'s are different, hence $\Phi({\mathbf k},\tau) \ne D\tau$ with a constant $D$ for all $k$'s.  Therefore, we can easily conclude that the waves are not advected by a constant mean velocity field, say ${\bf U}_0$.
\item The absolute values of the slopes  increase with $k$.  Note that the slopes come with both positive and negative signs. 
\end{enumerate}

Given the above properties, we postulate that for $\tau$ up to $ \tau_c$,
\be
 \Phi({\mathbf k},\tau) \sim {\bf k} \cdot \tilde{\bf U}_0({\bf k}) \tau \sim c k \tilde{U}_0(k) \tau, 
 \ee
where $c$ is a random number taking both positive and negative signs.      Hence the normalised correlation function as well as the Green's function need to be modified to
 \begin{eqnarray}
  R(\mathbf k, \tau) = G(\mathbf k, \tau) & = &  \exp(-\tau/\tau_c) \exp(-i {\bf k} \cdot \tilde{\bf U}_0({\bf k}) \tau) \nonumber \\
  & \sim &  \exp(-\tau/\tau_c) \exp(-i c k \tilde{U}_0(k) \tau), 
 \label{eq:Rk_tau_sweeping}
\end{eqnarray}
which is a numerical demonstration of the {\em sweeping effect} proposed by \citet{Kraichnan:PF1964Eulerian}.  Physically, a wave ${\bf u}({\bf k})$ is being advected by the {\em random mean velocity field},  $ \tilde{U}_0(k) $.   The random velocity changes its direction and magnitude in one eddy turnover time.  This is the reason why the phases are linear in $\tau$ only up to $\tau \approx \tau_c$.   The aforementioned wavenumber-dependent mean velocity field is in the similar spirit as the advection of eddies within eddies~\citep{Davidson:book:Turbulence,Pope:book,McComb:book:Turbulence}. 
 
Note that the  numerically-computed $C({\bf k},\tau)$ is of the form given by Eq.~(\ref{eq:Rk_tau_sweeping}), contrary to  Eq.~(\ref{eq:R_Kraichnan}), as argued by \citet{Kraichnan:PF1964Eulerian}.   This is an important deviation from earlier works on sweeping effect.  We also remark that it is important to investigate $k$-dependence of $\tilde{U}_0(k)$.   From scaling arguments, we expect $\tilde{U}_0(k)  \sim \epsilon^{1/3} k^{-1/3}$.  Unfortunately, the present simulation does not have sufficient resolution to test this conjecture. We need to perform a high-resolution simulation that will facilitate a larger range of $k$.  If the above conjecture is indeed correct, then the velocity at larger length scale would sweep smaller flow elements embedded within, thus validating the {\em eddies within eddies} viewpoint in turbulent flows.

From Eq.~(\ref{eq:Rk_tau_sweeping}), we deduce that in ${\bf k}-\omega$ space, the Green's function is
\be
G({\bf k},\omega) = \frac{1}{-i \omega + \nu(k) k^2 + i c k  \tilde{U}_0(k) }.
\label{eq:G_k_omega_new}
\ee
Note that the Eulerian field theory does not incorporate the additional term $i c k  \tilde{U}_0(k) $ of the above equation~\citep[see Appendix~\ref{sec:rgcalc} and][]{McComb:book:Turbulence,McComb:book:new,Mccomb:PRA1983,McComb:JFM1984,Zhou:PRA1988,Zhou:PR2010}. This is the reason why  \citet{Kraichnan:PF1964Eulerian} argued that the Eulerian field theory may not be appropriate for the description of hydrodynamic turbulence. To overcome this deficiency, \citet{Kraichnan:PF1965Lagrangian_history} devised a Lagrangian-based field-theoretic treatment of turbulence. Unfortunately, this topic is beyond the scope of this paper.

 Though field-theoretic treatment is not a main theme of the paper,  in Appendix~\ref{sec:rgcalc}, we discuss this topic very briefly.  In the Eulerian treatment of hydrodynamic turbulence, the Green's function is taken to be of the form of Eq.~(\ref{eq:Gk}).  The perturbative computation shows that the renormalized viscosity is independent of the mean velocity field ${\bf U}_0$, which may tempt us to believe that the Eulerian framework respects Galilean invariance.  However, as discussed in this section, a more realistic Green's function consistent with  the sweeping effect is of the form given by Eq.~(\ref{eq:G_k_omega_new}).  Unfortunately, incorporation of this Green's function makes the perturbative approach quite untenable, consistent with the arguments of \citet{Kraichnan:PF1964Eulerian} in which he claims nonsuitability of  Eulerian field theory for the field-theoretic treatment of turbulence.
 
 In the next section we will describe the frequency spectrum of turbulent flows with $U_0=0$.
 
\section{Frequency spectrum of $f^{-2}$ for turbulent flow}
\label{sec:f-2}
It is interesting to note that for  homogeneous and isotropic turbulence, the frequency spectrum $E(\omega)$ of the velocity time series measured by a real space probe is proportional to $\omega^{-2}$.  This spectrum can be deduced from the correlation function of Eq.~(\ref{eq:Rk_tau_sweeping}) in the following manner.   The correlation function $C({\bf r}, \tau)= \langle {\bf u(x},t) \cdot {\bf u(x+r},t+\tau) \rangle_{{\bf x},t} $, where $\langle . \rangle_{{\bf x},t}$ represents averaging over ${\bf x}$ and $t$, is
\bea
C({\bf r}, \tau) & = & \int d{\bf k} C({\bf k},\tau)  \exp{(i {\bf k \cdot r})}  \nonumber \\
 & = & \int d{\bf k} C({\bf k}) \exp(-\nu(k) k^2 \tau) \exp(-i {\bf k} \cdot \tilde{\bf U}_0({\bf k}) \tau)  \exp(i {\bf k} \cdot {\bf r}).
\eea 
Averaging the above for random $\tilde{\bf U}_0({\bf k})$ yields~\citep{Kraichnan:PF1964Eulerian,Wilczek:PRE2012}
\bea
C({\bf r}, \tau)  & = & \int d{\bf k} C({\bf k}) \exp(-\nu(k) k^2 \tau) \langle \exp(-i c k \tilde{U}_0( k)\tau)  \rangle \exp(i {\bf k} \cdot {\bf r}) \nonumber \\
 & = & \int d{\bf k} C({\bf k}) \exp(-\tau/\tau_c)  \exp(-k^2 [\tilde{U}_0( k)]^2 \tau^2)  \exp(i {\bf k} \cdot {\bf r}).
 \label{eq:Crtau}
\eea 
Here we replace the isotropic and homogeneous $C({\bf k})$ with 
\be
C({\bf k})  = \frac{E(k)}{4\pi k^2} = \frac{f_L(kL) f_\eta(k\eta) K_\mathrm{Ko} \epsilon^{2/3} k^{-5/3}}{4\pi k^2},
\label{eq:Ek_Pope}
\ee
where $\epsilon$ is the energy dissipation rate, same as the energy flux, and
\begin{eqnarray}
f_L(kL) & = & \left( \frac{kL}{[(kL)^2 + c_L]^{1/2}} \right)^{5/3+p_0}, 
\label{eq:fL} \\
f_\eta(k\eta) & = & \exp \left[ -\beta \left\{ [ (k \eta)^4 + c_\eta^4 ]^{1/4}   - c_\eta \right\} \right],
\label{eq:feta}
\end{eqnarray}
with $c_L, c_\eta, p_0, \beta$ as constants. Here $L$ is the large length scale. Note that, here we use \citet{Pope:book} model of a turbulent flow,
\begin{equation}
E(k) = K_\mathrm{Ko} \epsilon^{2/3} k^{-5/3}  f_L(k L)  f_\eta(k\eta), 
\label{eq:Ek_Pope1}
\end{equation}
to describe the energy spectrum $E(k)$. We also substitute $\tau_c(k) = 1/(\nu(k) k^2) = \epsilon^{-1/3}k^{-2/3}$ and $\tilde{U}_0(k) = \epsilon^{1/3}k^{-1/3}$ (as argued in the last section).  We ignore the coefficients in front of these quantities for brevity.  

Since we are measuring the velocity at  a single point, we set ${\bf r}=0$.  Hence the temporal correlation of the velocity field at a given point ${\bf x}$ is 
\bea
 C({\bf r}=0, \tau) & = & \int d{\bf k} C({\bf k}) \exp(-\nu(k) k^2 \tau )  \exp(-k^2 [\tilde{U}_0( k)]^2 \tau^2)  \nonumber \\
 & = &  K_\mathrm{Ko} \epsilon^{2/3} \int dk  k^{-5/3} f_L(kL) f_\eta(k\eta) \exp(-\epsilon^{1/3}k^{2/3}\tau ) \times  \nonumber \\
 	&&   \exp(-\epsilon^{2/3}k^{4/3}\tau^2).   
\eea 
Using $\tau(k) \sim 1/(k u_k) \sim \epsilon^{-1/3} k^{-2/3}$, we make a change of variable:
\be
k = \tilde{k} \epsilon^{-1/2} \tau^{-3/2},
\ee
that yields
\bea
 C(\tau) & = &  K_\mathrm{Ko} \epsilon  \tau \int d{\tilde{k}} \tilde{k}^{-5/3} f_L(\tilde{k} (L/U\tau)^{3/2}) f_\eta( \tilde{k} (\tau_d/ \tau)^{3/2})  \exp(- \tilde{k}^{2/3} - \tilde{k}^{4/3} ), \label{eq:C_tau_U_zero}
\eea 
where $U$ is the large-scale velocity, and $\tau_d$ is the dissipative time scale.  We focus on $\tau$  the inertial range, hence $L/U\tau \gg 1$ and $\tau_d/ \tau \ll 1$.  Therefore, from Eqs.~(\ref{eq:fL}),~(\ref{eq:feta}), $f_L(\tilde{k} (L/U\tau)^{3/2}) \approx 1$ and $f_\eta( \tilde{k} (\tau_d/ \tau)^{3/2}) \approx 1$ .     Therefore,
\bea
 C(\tau) =  K_\mathrm{Ko} \epsilon  \tau \int d{\tilde{k}} \tilde{k}^{-5/3} \exp(- \tilde{k}^{2/3} - \tilde{k}^{4/3} ) = A K_\mathrm{Ko} \epsilon  \tau,
\eea 
where $A$ is the value of the integral of Eq.~(\ref{eq:C_tau_U_zero}).  The Fourier transform of $ C(\tau)$ yields the frequency spectrum, which is
\bea
C(\omega) & = & \int C(\tau) \exp(i \omega \tau) d\tau = A K_\mathrm{Ko} \epsilon \int \tau  \exp(i \omega \tau) d\tau  \sim  \epsilon \omega^{-2} \sim  \epsilon f^{-2},
\label{eq:Eomega_minus2}
\eea
where $f=\omega/(2\pi)$. Thus we show that the frequency spectrum $E(f) \sim f^{-2}$. 

\begin{figure}
\begin{center}
\includegraphics[scale = 0.75]{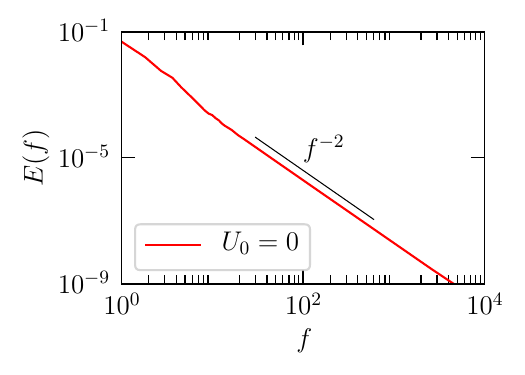}
\end{center}
\caption{For $U_0=0$, plot of the frequency spectrum $E(f)$ for the velocity time series measured by  a real space probe.  The plot is averaged over 50 real space probes located at random locations.  Here $E(f) \sim f^{-2}$, consistent with the sweeping effect  (see Eq.~(\ref{eq:Eomega_minus2})).}
\label{fig:freq_spec_U0}
\end{figure}

To compute the frequency spectrum $E(f)$ from the time series of the velocity field, we placed $50$ real space probes at  random locations in the cubical box.  We record the three components of the velocity field at all the real space probes. We run our simulation for single eddy turnover time with a constant $\Delta t=3\times10^{-5}$, which helps us compute the Fourier transform of the real space data using equispaced FFT.  Note that for $\Delta t=3\times10^{-5}$, the Courant number is less than unity; hence our simulation is well resolved in time. We record the velocity fields at every $33$ steps; thus we have $10^3$ data points. Then we perform Fourier transform of the velocity components $u_i(t)$ ($i=x,y,z$) and compute the frequency spectrum $E(f)$ using following formula
\begin{equation}
E(f) = \frac{1}{2} \left( |\hat{u}_x(f)|^2 + |\hat{u}_y(f)|^2 +  |\hat{u}_z(f)|^2 \right).
\label{eq:Ef}
\end{equation}
Figure~\ref{fig:freq_spec_U0} exhibits the averaged frequency spectrum that exhibits $f^{-2}$ scaling, consistent with  Eq.~(\ref{eq:Eomega_minus2}).    Earlier, \citet{Landau:book:Fluid} had derived the aforementioned power law using dimensional analysis.  We will revisit their arguments in the next section.

 In the next section, we will report the numerical results for ${\bf U}_0 = 10\hat{z}$  and show the effects of the mean velocity field on the correlation function, which is related to  Taylor's frozen-in turbulence hypothesis.

\section{Taylor's frozen-in turbulence hypothesis}
\label{sec:Taylor}
We compute the correlation function when $U_0 \ne 0$; this phenomenon is related to Taylor's frozen-in turbulence hypothesis.  In this paper, we present our results for ${\bf U}_0 = 10\hat{z}$.

As described in \S\ref{sec:sweeping_effect}, we performed numerical simulation of incompressible fluid with ${\bf U}_0 = 0$ and ${\bf U}_0 = 10\hat{z}$.  Compared to the flow for $ U_0=0$, the flow profile for ${\bf U}_0 = 10\hat{z}$ is shifted upward as shown in Fig.~\ref{fig:real_space}, and the energy spectrum and flux are the same as  for $U_0=0$ and 10.   The correlation function for ${\bf U}_0 = 10\hat{z}$, however, exhibits features very different from that for $U_0=0$; this phenomenon is related to Taylor's hypothesis.

\begin{figure}
\begin{center}
\includegraphics[scale = 0.7]{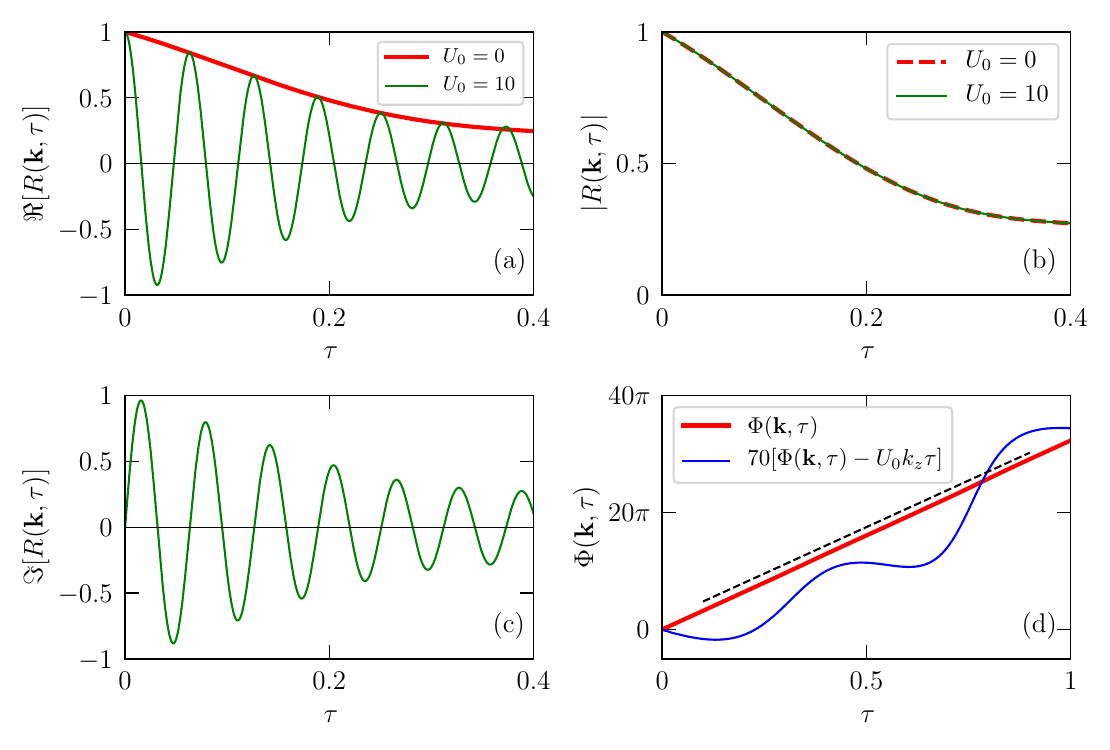}
\end{center}
\caption{For $\mathbf U_0 = 10$ and  $\mathbf k = (0,0,10)$ in the inertial range), plots of the normalised correlation function $R(\mathbf k, \tau)$ vs.~$\tau$:  (a) $\Re[R(\mathbf k, \tau)]$, (b) $|R(\mathbf k, \tau)|$, (c) $\Im[R(\mathbf k, \tau)]$, and (d) $\Phi(\mathbf k, \tau)$.  The real and imaginary parts exhibit damped oscillation with the frequency of $U_0 k$ and damping time of $1/(\nu(k) k^2)$.  $|R(\mathbf k, \tau)|$ for $U_0=0,10$ are identical, thus showing that the decay time scales for the two cases are the same; also, $|R(\mathbf k, \tau)|$ provides envelop to the real part.  The phase of  $R(\mathbf k, \tau)$ varies as $\Phi({\bf k},\tau) = U_0 k_z \tau + \delta$, where  $\delta$  arises due to the sweeping by the random large-scale flow structures.  The dashed black and  blue lines represent $U_0 k_z \tau$ and $70 \delta$ (amplified by a factor for visualisation) respectively.  }
\label{fig:U10}
\end{figure}

We compute $R(k,\tau)$ of Eq.~(\ref{eq:R}) using the numerical data.  We choose ${\bf k} = (0,0,10)$. In Fig.~\ref{fig:U10}, we plot  the real and imaginary parts of the correlation $R(\mathbf k, \tau)$, as well as its magnitude and phase.  As shown in the figure, $|R(\mathbf k, \tau)|$ is approximately same for $U_0=0$ and 10.  However, both the real and imaginary parts of $R(\mathbf k, \tau)$ exhibit damped oscillations with a frequency of $\omega = k_z U_0$ and a decay time scale of $1/(\nu(k) k^2)$.  This is evident from the envelop of $\Re[R(\mathbf k, \tau)]$ that matches  with $\Re[R(\mathbf k, \tau)]$ for $U_0=0$.   Hence, we may naively expect that
\begin{equation}
  R(\mathbf k, \tau) = G(\mathbf k, \tau)= \exp(-\tau/\tau_c) \exp(i {\bf U}_0 \cdot {\bf k} \tau).
 \label{eq:Rk_tau_sweeping1}
\end{equation}
However, there are some signatures of {\em random sweeping effect} for $U_0=10$ as well.  In Fig.~\ref{fig:U10}, we plot the phase $\Phi$ of $R(\mathbf k, \tau)$, which is quite close to $U_0 k \tau$.  However, $\Phi - U_0 k \tau$ is nonzero, which is evident  from its magnified plot in Fig.~\ref{fig:U10}(d).  This deviation is due to the random sweeping effect by random mean field $\tilde{U}_0$. Thus, the small-scale fluctuations are swept by ${\bf U}_0=10 \hat{z}$ and by large-scale random velocity $\tilde{U}_0(k)$.   The effects of $\tilde{U}_0(k)$ for ${\bf U}_0=10 \hat{z}$ and ${U}_0=0$ are expected to be the same, since the velocity fluctuations are the same in both the flows. Therefore, the Green's function and the normalised correlation function for ${\bf U}_0=10 \hat{z}$ can be written as 
 \begin{equation}
  R(\mathbf k, \tau) = G(\mathbf k, \tau)= \exp(-\tau/\tau_c) \exp(-i {\bf U}_0 \cdot {\bf k} \tau - c k \tilde{U}_0(k) \tau),
 \label{eq:Rk_tau_Unonzero}
\end{equation}
and
\be
G({\bf k},\omega) = \frac{1}{-i \omega + \nu(k) k^2 + i {\bf U}_0 \cdot {\bf k} +  i c k  \tilde{U}_0(k) }.
\label{eq:G_k_omega_new_Unonzero}
\ee

Now let us discuss Taylor's frozen-in turbulence hypothesis, according to which the frequency spectrum of the real-space velocity time series $E(f) \sim f^{-5/3}$.  This conclusion can be easily derived using the correlation function of Eq.~(\ref{eq:Rk_tau_Unonzero}).  Following the same set of mathematical steps as in the earlier section on the correlation function of Eq.~(\ref{eq:Rk_tau_Unonzero}), we obtain
\bea
 C({\bf r}, \tau) & = & \int d{\bf k} C({\bf k}) \exp(-\nu(k) k^2 \tau-i{\bf U_0 \cdot k} \tau)  \exp(-i {\bf k} \cdot \tilde{\bf U}_0({\bf k}) \tau)  \exp(i {\bf k} \cdot {\bf r}).
 \label{eq:corr_U0}
\eea 
We time average $\tilde{U}_0$ over random ensemble that yields
\bea
 C({\bf r}, \tau) & = & \int d{\bf k} C({\bf k}) \exp(-\nu(k) k^2 \tau - i{\bf U_0 \cdot k} \tau) \langle \exp(-i c k \tilde{U}_0( k)\tau)  \rangle \exp(i {\bf k} \cdot {\bf r}) \nonumber \\
 & = & \int d{\bf k} C({\bf k}) \exp(-\tau/\tau_c - i{\bf U_0 \cdot k} \tau )  \exp(-k^2 [\tilde{U}_0( k)]^2 \tau^2)  \exp(i {\bf k} \cdot {\bf r}).
\eea 
Since we are measuring the velocity at  a single point, we  set ${\bf r}=0$.  We also replace $C({\bf k})$ of the above equation with that of Eq.~(\ref{eq:Ek_Pope}) that yields  
\bea
 C(\tau) & = &  K_\mathrm{Ko} \epsilon^{2/3} \int dk  k^{-5/3} f_L(kL) f_\eta(k\eta) \exp(-i{\bf U_0 \cdot k} \tau)  \times  \nonumber \\
 	&&  \exp(-\epsilon^{1/3}k^{2/3}\tau )  \exp(-\epsilon^{2/3}k^{4/3}\tau^2).
\eea 
For the integration, we choose the $z$ axis along the direction of ${\bf U}_0$. Since $\tau \sim 1/(U_0 k)$ is the dominant time scale, we  make a change of variable:
\be
\tilde{k} =   U_0 k \tau
\ee
that yields
\bea
 C(\tau) & = &  K_\mathrm{Ko}  (\epsilon U_0 \tau)^{2/3}  \int d{\tilde{k}} \tilde{k}^{-5/3} f_L(\tilde{k} (L/U_0\tau)) f_\eta( \tilde{k} (\eta / U_0 \tau)  \frac{\sin(U_0 k \tau)}{U_0 k \tau} \times   \nonumber \\
 &&  \exp[- \tilde{k}^{2/3} (U/U_0)^{2/3} (\tau/T)^{1/3} - \tilde{k}^{4/3} (U/U_0)^{4/3} (\tau/T)^{2/3} ].
\eea 
We focus on $\tau$ in the inertial range, hence $L/U_0\tau \gg 1$ and $\eta/ U_0 \tau \ll 1$,  consequently,  $f_L(\tilde{k} (L/U_0\tau)) \approx 1$, and $f_\eta( \tilde{k} (\eta / U_0 \tau) \approx 1$.     Therefore,
\bea
 C(\tau) & = & K_\mathrm{Ko} (\epsilon U_0 \tau)^{2/3}   \int d{\tilde{k}} \tilde{k}^{-5/3} \frac{\sin \tilde{k}}{\tilde{k} }  \exp[- \tilde{k}^{2/3} (U/U_0)^{2/3} (\tau/T)^{1/3} - \tilde{k}^{4/3} (U/U_0)^{4/3} (\tau/T)^{2/3} ] \nonumber \\ 
 & = & B \alpha K_\mathrm{Ko}(\epsilon U_0 \tau)^{2/3},
\eea 
where $B$ is the value of the integral.  The Fourier transform of the above $ C(\tau)$ yields the frequency spectrum
\bea
C(\omega) & = & \int C(\tau) \exp(i \omega \tau) d\tau = \int  B K_\mathrm{Ko}(\epsilon U_0 \tau)^{2/3} \exp(i \omega \tau) d\tau  \nonumber \\
& \sim  & (\epsilon U_0)^{2/3} \omega^{-5/3} \sim (\epsilon U_0)^{2/3} f^{-5/3}.
\eea
This is the Taylor's  frozen-in turbulence hypothesis~\citep{Taylor:PRS1938}, according to which the frequency spectrum of the velocity field measured by a real space probe also yields Kolmogorov's $-5/3$ spectrum.  This is very useful hypothesis because to determine $E(k)$, we do not need to measure the velocity field at all physical locations by expensive experimental setups.  Researchers have exploited the above hypothesis to measure turbulence spectrum in many fluid and plasma experiments, for example in wind tunnels, and measurement of solar wind turbulence using extraterrestrial spacecrafts.

We could also argue the above frequency spectrum using scaling arguments~\citep{Landau:book:Fluid}.  From the definition of Green's function~(\ref{eq:G_k_omega_new_Unonzero}), we obtain the dominant $\omega = \mathbf U_0 \cdot \mathbf k +k \tilde{U}_0(k)  -i \nu(k) k^2$.  When $\mathbf U_0 \cdot \mathbf k \gg \nu(k) k^2$ and $\mathbf U_0 \cdot \mathbf k \gg k \tilde{U}_0(k)$, we obtain $\omega = U_0 k_z$.  Therefore, using  $E(k) = K_\mathrm{Ko} \Pi^{2/3} k^{-5/3}$, we obtain
\begin{eqnarray}
E(\omega) & = &  E(k) \frac{dk}{d\omega}  
\sim K_\mathrm{Ko} \Pi^{2/3} (\omega/U_0)^{-5/3} (1/U_0) \nonumber \\
& \sim & K_\mathrm{Ko}  (U_0\Pi)^{2/3} \omega^{-5/3},
\label{eq:f5by3scaling}
\end{eqnarray}
consistent with the principle of Taylor's frozen-in turbulence hypothesis.  Here we have replaced $k$ by $k_z$.  On the contrary, when $\mathbf U_0 \cdot \mathbf k \ll\nu(k) k^2$ (for zero or small $U_0$), we obtain 
\begin{equation}
\omega \approx  \nu(k) k^2 = \nu_* \sqrt{K_\mathrm{Ko}} \Pi^{1/3} k^{2/3}, 
\end{equation}
and hence, using $E(k) = K_\mathrm{Ko} \Pi^{2/3} k^{-5/3}$, we obtain
\begin{eqnarray}
E(\omega) & = & E(k) \frac{dk}{d\omega} \nonumber \\
& = & \frac{K_\mathrm{Ko} \Pi^{2/3} k^{-5/3}}{\nu_* \sqrt{K_\mathrm{Ko}} \Pi^{1/3} (2/3) k^{-1/3}} \nonumber \\
& = & \frac{3}{2} \nu_* (K_\mathrm{Ko})^{3/2} \Pi \omega^{-2},
\label{eq:f2scaling}
\end{eqnarray}
as derived by \citet{Landau:book:Fluid}.  Thus, the Green's function of Eq.~(\ref{eq:G_k_omega_new_Unonzero}) helps us deduce both $\omega^{-5/3}$ and $\omega^{-2}$ frequency spectra depending on the strength of ${\bf U}_0$.  The above discussion also demonstrates that the Eulerian picture picks up both,  the sweeping effect and Taylor's hypothesis.  We refer the reader to \citet{Tennekes:JFM1975} and \citet{He:ARFM2016} for Eulerian and Lagrangian time scales, and their connection to $\omega^{-2}$ and $\omega^{-5/3}$ power spectra.

\begin{figure}
\begin{center}
\includegraphics[scale = 0.7]{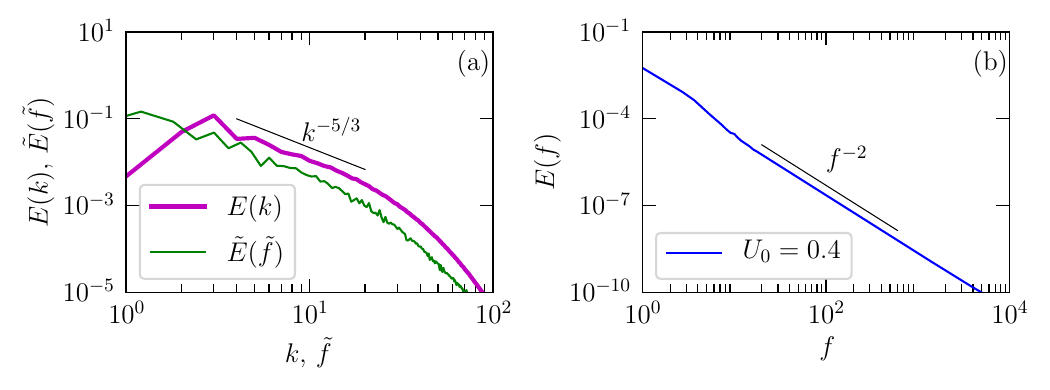}
\end{center}
\caption{(a) For $U_0=10$, plots of the wavenumber spectrum $E(k)$ and the scaled frequency spectrum $E(f)$ for the velocity time series measured by  real-space probes.  The plot is averaged over 50 real-space probes located at random locations.  Here $\tilde{f} = f (2\pi)/U_0$ and $\tilde{E}(\tilde{f}) = E(f) U_0/(2\pi)$.  $E(f) \sim f^{-5/3}$, consistent with Taylor's frozen-in turbulence hypothesis.  (b) For $U_0=0.4$, $U_0 \ll \tilde{U}_0$, hence $E(f) \sim f^{-2}$, consistent with the sweeping effect. Also see Fig.~\ref{fig:freq_spec_U0}.}
\label{fig:freq_spec}
\end{figure}

To test Taylor's  frozen-in turbulence hypothesis numerically, we record time series of the velocity field at the real space probes, and then compute their  frequency spectrum $E(f)$, same analysis as that of the previous section, but here with ${\bf U}_0 = 10\hat{z}$. Figure~\ref{fig:freq_spec}(a) exhibits  $E(k)$ and $E(f)$  in the presence of mean velocity field ${\bf U}_0 = 10\hat{z}$, for which both the spectra follow Kolmogorov's scaling (see Eq.~(\ref{eq:f5by3scaling})). Note that in Fig.~\ref{fig:freq_spec}(a) we scaled the frequency spectrum: $f \rightarrow \tilde{f} = f (2\pi)/U_0$ and $E(f) \rightarrow \tilde{E}(\tilde{f}) = E(f) U_0/(2\pi)$. Our result that $\tilde{E}(\tilde{f}) \approx E(k)$ illustrates Taylor's  hypothesis in the presence of a mean velocity field.  


For $U_0 \ll u_L$ or $U_0 \ll \epsilon^{1/3} L^{1/3}$, as argued above, Taylor's frozen-in turbulence hypothesis will not work.  Rather, the sweeping effect by local mean velocity would dominate the dynamics, hence we expect $E(f) \sim f^{-2}$.  To test this hypothesis, we perform a numerical simulation for ${\bf U}_0 = 0.4\hat{z}$ following the same procedure as that for ${\bf U}_0 = 10\hat{z}$.   For  ${\bf U}_0 = 0.4\hat{z}$, Fig.~\ref{fig:freq_spec}(b) exhibits the averaged frequency spectra, which exhibit $f^{-2}$ scaling, which is consistent with the above arguments.

In the next section, we describe elliptic approximation that relates space-time correlation to equal-time correlation function.

\section{Elliptic approximation}
\label{sec:elliptic}
Recently \citet{He:PRE2010}, \citet{He:PRE2011}, and \citet{He:ARFM2016} attempted to combine  the sweeping effect with Taylor's frozen-in turbulence hypothesis.  Here we reproduce their arguments using Eq.~(\ref{eq:corr_U0}).

\begin{figure}
\begin{center}
\includegraphics[scale = 0.9]{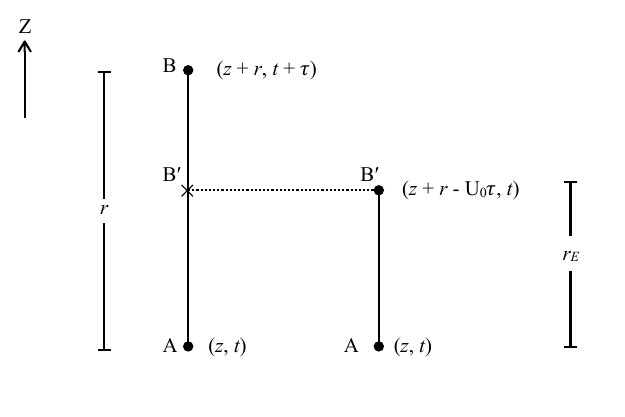}
\end{center}
\caption{A and B represent respectively the velocity measurements at locations $z$ and $z+r$ and at times $t$ and $t+\tau$. The fluid element at B would be at ${\rm B}^\prime$ at time $t$, thus A and ${\rm B}^\prime$ would represent equal-time measurements.  Note that $r_E = r-U_0 \tau$. }
\label{fig:elliptic}
\end{figure}

We consider fluid flow with a mean velocity of $U_0$ along the $z$ axis.   We focus on  the vertical velocities measured at two points $z$ and $z+r$, but at  times $t$ and $t+\tau$ (see Fig.~\ref{fig:elliptic} for an illustration).   For the same, the space-time correlation derived using Eq.~(\ref{eq:corr_U0}) is
\bea
 C(r, \tau) & = & \int d{\bf k} C({\bf k}) \exp(-\nu(k) k^2 \tau) \exp\left\{ [r -i(U_0 +\tilde{U}_{0z}) \tau)]  i  k_z -i  \tilde{\bf U}_{0\perp} \cdot {\bf k}_\perp \tau \right\}.
\eea 
Now suppose that 
\be
r \approx U_0 \tau  \gg  \nu(k) k^2 \tau,
\ee
then
\be
C(r,\tau) =  \int d{\bf k} C({\bf k})    \exp\left\{ [r -(U_0 +\tilde{U}_{0z}) \tau)]  i  k_z -i  \tilde{\bf U}_{0\perp} \cdot {\bf k}_\perp \tau \right\}.  
\label{eq:Crtau_elliptic}
\ee
We can relate the above correlation function to an equal-time correlation function
\be
C({\bf r}_E, 0)  =  \exp[ i r_{Ez} k_z + i {\bf r}_{E\perp} \cdot {\bf k}_{\perp}]
\label{eq:C_rE}
\ee
with
\be
r_{Ez} =  [r -(U_0 +\tilde{U}_{0z}) \tau)];~~ {\bf r}_{E\perp} = \tilde{\bf U}_{0\perp} \tau
\ee
or
 \begin{equation}
r_E^2 = r_{Ez}^2 +   |{\bf r}_{E\perp}|^2 =  (r-U\tau)^2 + (V\tau)^2,
\label{eq:rE}
\end{equation}
where
\bea
U & = &  U_0 +\tilde{U}_{0z}  \\
V & = & |\tilde{\bf U}_{0\perp}|.
\eea
This is the statement of elliptic approximation~\citep{He:PRE2010,He:PRE2011,He:ARFM2016}. Our derivation is slightly different from those of \citet{He:PRE2010}, \citet{He:PRE2011}, and \citet{He:ARFM2016}.  

Thus, the elliptic approximation includes both, the sweeping effect and Taylor's frozen-in turbulence hypothesis.  The velocities $U_0$ and $\tilde{U}_0$ yield the Eulerian and Lagrangian space-time correlations respectively, and they are related to the sweeping effect and Taylor's hypothesis respectively.  It is easy to see  that the conventional Taylor's hypothesis is applicable when $U_0 \gg \tilde{U}_0$ and it yields $f^{-5/3}$ spectrum, for which the physical interpretation is as follows.   The velocity correlation for the velocity measurements at A and B of Fig.~\ref{fig:elliptic}, $ C(r, \tau) = \langle {\bf u}(z,t) {\bf u}(z+r,t+\tau) \rangle$, is same  as those measured at A and ${\rm B}^\prime$ at the same time $t$, $ C(r_E, 0) = \langle {\bf u}(z,t) {\bf u}(z+r- U_0 \tau,t) \rangle$.  This is because the fluid element at ${\rm B}^\prime$ at time $t$ reaches B at time $t+\tau$.  Note that Taylor's frozen-in turbulence hypothesis assumes that $r=0$; for this case, ${\rm B}^\prime$ would be at $z-U_0 \tau$.


\citet{Sreenivasan:PRL1996Shear} analysed the velocity fluctuations of the atmospheric data and showed that the conditional expectation of $\Delta u_r^2$ depends on the  {\em local mean velocity field} $u_0$ for small Reynolds number, but it is independent of $u_0$ for large Reynolds number.  Here  $\Delta u_r = u(x+r)-u(x)$, where $u$ and $r$ are respectively the velocity component and the separation distance in the direction $x$. This is due to the dominance of Lagrangian space-time correlation~\citep{He:ARFM2016}.  In a related work, \citet{Cholemari:JFM2006} proposed a model to relate the spatial and temporal Eulerian two-point correlations in the absence of mean flow.  We also remark that \citet{He:PRE2010} employed elliptic approximation to Rayleigh-B\'{e}nard convection, and related the frequency spectrum of a real-space probe to the energy spectrum, $E(k)$.   

\section{Discussions and Conclusions}
\label{sec:conclusion}

Using numerical simulations, we investigate the sweeping effect and Taylor's frozen-in turbulence hypothesis, and show consistency between them.  We performed numerical simulations with and without mean flow ($U_0=10$ and 0 respectively).   The velocity fluctuations for the two cases exhibit identical energy spectra and energy fluxes, but the space-time correlations for the two cases are different.

For  $U_0=0$, we compute the velocity correlation function $C(k,\tau)$ and show that its real part decays with time-scale $1/(\nu(k) k^2)$, where $\nu(k)$ is the renormalised viscosity.  However, the phase of the correlation function shows a linear increase with $\tau$ till approximately one eddy turnover time; this is attributed to the sweeping  by {\em random mean velocity} of the flow.  Thus we demonstrate a clear signature of sweeping effect in hydrodynamic turbulence.  Our approach deviates from those of \citet{Sanada:PF1992} who use absolute of correlation function.  Note that the phase of the correlation function extracts the effects of the sweeping effect by random mean velocity.

For $U_0=10$, the correlation function exhibits damped oscillations with a frequency of $\omega=U_0 k$ and decay time scale of $1/(\nu(k) k^2)$; the decay time scales for $U_0=10$ is same as that for $U_0=0$. A careful examination of the  phase of the correlation function also shows additional variations due to random velocity of the flow.

For the aforementioned two cases, the frequency spectra of the velocity field measured by real-space probes are different.  For $U_0=0$, $E(f) \sim f^{-2}$, which is related to the Lagrangian space-time correlation, but for $U_0=10$, $E(f) \sim f^{-5/3}$, which is the predictions of Taylor's frozen-in turbulence  hypothesis.  We demonstrate these spectra from their respective space-time correlation functions.  Our analysis shows that Taylor's hypothesis is applicable when
\be
U_0 k \gg \nu(k) k^2;~~~U_0 \gg \tilde{U}_0,
\ee
where $\tilde{U}_0$ is random mean velocity, which is responsible for the sweeping effect.

Thus, we provide a systematic demonstration of sweeping effect and Taylor's  frozen-in turbulence hypothesis, and show consistency between the two contrasting phenomena.  We demonstrate the above spectra using numerical simulations.

\section*{Acknowledgements}
We thank Sagar Chakraborty, K. R. Sreenivasan, Robert Rubinstein, Victor Yakhot, and Jayanta K. Bhattacharjee for useful discussions and suggestions.  Our numerical simulations were performed on {\em Chaos} clusters of IIT Kanpur. This work was supported by the research grants PLANEX/PHY/2015239 from Indian Space Research Organisation, India, and  by the  Department of Science and Technology, India (INT/RUS/RSF/P-03) and Russian Science Foundation Russia (RSF-16-41-02012) for the Indo-Russian project.

\appendix
\section{Renormalization group analysis in the presence of $\mathbf U_0$}
\label{sec:rgcalc}
 Hydrodynamic turbulence involves multi scales, hence, renormalization group (RG) formulation is a useful tool to  study hydrodynamic turbulence.  Some of the leading efforts on the RG formulation of hydrodynamic turbulence are: Yakhot-Orszag perturbative approach~\citep{Yakhot:JSC1986}, self-consistent approach of \citet{McComb:book:Turbulence} and \citet{Zhou:PR2010}, and generating functional formulation of \citet{DeDominicis:PRA1979}.  See recent review of \citet{Zhou:PR2010} for further discussion.  Most of the aforementioned computations are for zero mean flow.  In this Appendix we make an extension of McComb's procedure~\citep{McComb:book:Turbulence} for $U_0\ne 0$.

 Navier--Stokes equations describe fluid flows in real space.  The corresponding equations in the Fourier space are 
\begin{eqnarray}
(-i\omega+  i {\mathbf U_0   \cdot \mathbf k}  +   \nu k^{2} )u_{i}(\hat{{k}})  & = &   -\frac{i}{2}P_{ijm}(\mathbf{k})\int_{\hat{{p}}+\hat{{q}}=\hat{{k}}}d\hat{{p}}\left[u_{j}(\hat{{p}})u_{m}(\hat{{q}})\right] + f_i(\hat{k}) ,\label{eq:udot}  \\
k_i u_i({\bf k}) & = & 0,
\end{eqnarray}
where
\begin{eqnarray}
P_{ijm}(\mathbf{k})  =  k_{j}P_{im}(\mathbf{k})+k_{m}P_{ij}(\mathbf{k}),\\
\label{eq:Pp}
\hat{k} = (\omega, \mathbf k), \hat{p} = (\omega', \mathbf p), \mathrm{and} \, \, \hat{q} = (\omega'', \mathbf q). \nonumber
\end{eqnarray}
 We  compute the renormalized viscosity in the presence of a mean velocity $\mathbf U_0$.
In this renormalization process, the wavenumber range $(k_{N},k_{0})$ is divided logarithmically into $N$ shells. The $n$th shell is $(k_{n},k_{n-1})$ where $k_{n}=h^{n}k_{0}\,\,(h<1)$ and $k_N = h^N k_0$. In the first step,  the spectral space is divided in two parts: the shell $(k_{1},k_{0})=k^{>}$, which is to be eliminated, and  $(k_{N},k_{1})=k^{<}$, set of modes to be retained. The   equation  for a Fourier mode $\hat{k}$ belonging to $k^{<}$ is
\begin{eqnarray}
\bigl[ -i\omega + &  & i {\mathbf U_0 \cdot \mathbf k} +  \nu_{(0)} k^2  \bigr]  u_{i}^{<}(\hat{k})  =   -\frac{i}{2}P_{ijm}({\textbf{k}}) \int_{\hat{{p}}+\hat{{q}} = \hat{k}}  d\hat{p}([u_{j}^{<}(\hat{p})u_{m}^{<}(\hat{q})]\nonumber \\
 &  & +2[u_{j}^{<}(\hat{p})u_{m}^{>}(\hat{q})]+[u_{j}^{>}(\hat{p})u_{m}^{>}(\hat{q})]) + f^<_i(\hat{k}),
\label{eq:ukless}
\end{eqnarray}
where $\nu_{(0)} = \nu$.  The equation for $u_{i}^{>}(\hat{k})$ modes can be obtained by interchanging $<$ and $>$ in the above  equations. 

The objective of the renormalization group procedure is to compute the corrections to the viscosity, $\delta \nu_{(0)}$, due to the second and third terms in the RHS of Eq.~(\ref{eq:ukless}).  The steps shown below are same as  the i-RG or iterative averaging RG procedure~\citep{McComb:book:Turbulence,McComb:book:new,Mccomb:PRA1983,McComb:JFM1984,Zhou:PRA1988,Zhou:PR2010,Verma:PR2004}.
\begin{enumerate}
\item The terms given in the second and third brackets in the right-hand side of Eq.~(\ref{eq:ukless}) are computed perturbatively. Since we are interested in the statistical properties of the velocity fluctuations, we perform an  ensemble average of the system~\citep{Yakhot:JSC1986}. It is assumed that $\mathbf{u}^{>}(\hat{k})$
have a gaussian distribution with a zero mean, while $\mathbf{u}^<(\hat{k})$  is unaffected by the averaging process. Hence,
\begin{eqnarray}
\left\langle u_{i}^{>}(\hat{k})\right\rangle  = 0;~~ \left\langle u_{i}^{<}(\hat{k})\right\rangle  = u_{i}^{<}(\hat{k}).
\label{eqn:avgend}
\end{eqnarray}
The homogeneity of turbulent fluctuations yields~\citep{Batchelor:book:Turbulence}
\begin{eqnarray}
\left\langle u_{i}^{>}(\hat{p})u_{j}^{>}(\hat{q})\right\rangle  & = & P_{ij}(\mathbf{p})C(\hat{p})\delta(\hat{p}+\hat{q}).\label{eq:nonhelical-uu}
\end{eqnarray}
The triple order correlations $\left\langle u_{i}^{>}(\hat{k})u_{j}^{>}(\hat{p})u_{m}^{>}(\hat{q})\right\rangle $ are zero due to the Gaussian nature of the fluctuations.  In addition, we  neglect the contribution from the triple nonlinearity $\left\langle u^{<}(\hat{k})u_{j}^{<}(\hat{p})u_{m}^{<}(\hat{q})\right\rangle $, as assumed in some of the  RG calculations of turbulence~\citep{Yakhot:JSC1986,Zhou:PRA1988,McComb:book:Turbulence,Zhou:PR2010,McComb:book:new}. The effects of triple nonlinearity can be included following the scheme proposed by \citet{Zhou:PRA1988}  and \citet{Zhou:PR2010}.

\item To  first order, the second bracketed term of  Eq.~(\ref{eq:ukless}) vanishes, but the nonvanishing third bracketed term yields corrections to $\nu_{(0)}$~\citep{McComb:book:Turbulence,McComb:book:new,Mccomb:PRA1983,McComb:JFM1984,Zhou:PRA1988,Zhou:PR2010,Verma:PR2004}.  Consequently, Eq.~(\ref{eq:ukless}) becomes
 \begin{eqnarray}
 \bigl[ -i\omega && +  i  {\mathbf U_0 \cdot \mathbf k} + ( \nu_{(0)}(k) +\delta\nu_{(0)}(k)) k^2 \bigr] u_{i}^{<}(\hat{k})  = \nonumber \\
&& -\frac{i}{2}P_{ijm}({\textbf{k}}) \int_{\hat{{p}}+\hat{{q}} = \hat{k}} \frac{d \mathbf p d\omega'}{(2\pi)^{d+1}} [u_{j}^{<}(\hat{p})u_{m}^{<}(\hat{k}-\hat{p})] + f^<_i(\hat{k} )
\label{eq:renormalized}
\end{eqnarray}
 with 
\begin{eqnarray}
\delta \nu_{(0)}(\hat{k}) k^2  = \frac{1}{d-1}\int_{\hat{p}+\hat{q}=\hat{k}}^{\Delta}  \frac{d\mathbf p d\omega'}{(2\pi)^{d+1}} [B(k,p,q)G(\hat{q})C(\hat{p})],  
\label{eq:renormalized_GC}
\end{eqnarray}
where 
\begin{equation}
B(k,p,q)=k p [(d-3)z+2z^{3}+(d-1)x y]  \label{eq:Bkpq}
\end{equation}
with $d$ is the space dimensionality,  $x,y,z$ are the direction cosines of ${{\mathbf k}, {\mathbf p}, {\mathbf q}}$, and  the Green's function $G(\hat{q}) $ is defined as
\begin{equation}
G(\hat{q}) = \frac{1}{-i \omega'' + i {\mathbf U_0 \cdot \mathbf q} + \nu_{(0)}(q) q^2}.
\label{eq:Green_fn_RG}
\end{equation}
It is assumed in the RG calculation of turbulence that the correlation function and  the Green's function have the same frequency dependence,  which is a generalization of fluctuation dissipation theorem~\citep{McComb:book:Turbulence}. Hence,  the correlation function $C(\hat{p})$ is defined as
\begin{equation}
C(\hat{p}) = \frac{C({\bf p})}{-i \omega' + i {\mathbf U_0 \cdot \mathbf p} + \nu_{(0)}(p) p^2},
\label{eq:correlation_fn_RG}
\end{equation}
where $C({\bf p})$ is the modal energy spectrum.  In \S\ref{sec:Taylor}, we show that the sweeping effect induces an additional term of the form $ i {\tilde{\mathbf U}_0 \cdot \mathbf p}$ in Green's function (see Eq.~(\ref{eq:G_k_omega_new_Unonzero})). 

\item A substitution of  Green's function and the correlation function in Eq.~(\ref{eq:renormalized_GC}) yields

\begin{eqnarray}
 \delta \nu_{(0)}(\hat{k}) k^2 =   \frac{1}{d-1} && \int_{\hat{p}+\hat{q}=\hat{k}}  \frac{d\mathbf p d\omega'}{(2\pi)^{d+1}} B(k,p,q) C(\mathbf p) \nonumber \\ \nonumber && \times
  \frac{  1}{[-i\omega'' + i {\mathbf U_0 \cdot \mathbf q} + \nu_{(0)}(q) q^2]} \\ &&  \times \frac{1}{ [-i\omega' + i {\mathbf U_0 \cdot \mathbf p} + \nu_{(0)}(p) p^2]} .
\label{eq:delta_nu}
\end{eqnarray}

Using  $\omega = \omega' + \omega''$, we obtain
\begin{eqnarray}
 \delta \nu_{(0)}(\omega, k)  k^2   & = &    \frac{1}{d-1} \int_{\hat{p}+\hat{q}=\hat{k}}  \frac{d\mathbf p d\omega'}{(2\pi)^{d+1}}  B(k,p,q)  C(\mathbf p) \nonumber \\ \nonumber &&  \times \frac{1}{\bigl[ -i\omega + i\omega'  + i {\mathbf U_0 \cdot \mathbf q} + \nu_{(0)}(q) q^2 \bigr]  \bigl[-i\omega' + i {\mathbf U_0 \cdot \mathbf p} + \nu_{(0)}(p)p^2 \bigr]} \nonumber \\
 &  = &   \frac{1}{d-1} \int_{\bf {p + q = k}}^{\Delta}  \frac{d\mathbf p}{(2\pi)^d}  B(k,p,q)  C(\mathbf p) \nonumber  \\  &&  \times  \frac{1}{ \bigl[ -i\omega  + \nu_{(0)}(p) p^2 + \nu_{(0)}(q) q^2 + (   i {\mathbf U_0 \cdot \mathbf p} + i {\mathbf U_0 \cdot \mathbf q})  \bigr]} \nonumber \\ 
 &   = &  \frac{1}{d-1}  \int_{\bf {p + q = k}}^{\Delta}  \frac{d\mathbf p}{(2\pi)^d}
 B(k,p,q)  C(\mathbf p)  \nonumber \\ && \times  \frac{1}{ \bigr[ -i(\omega - {\mathbf U}_0 \cdot {\mathbf k}) + \nu_{(0)}(p) p^2 + \nu_{(0)}(q) q^2 \bigr]}.
\label{eq:delta_nu2}
\end{eqnarray}

We employ a contour integral to integrate $\omega'$ to go from the first step to the second step of Eq.~(\ref{eq:delta_nu2}).  The integration $d\bf p$ is performed over the wavenumber shell $(k_1,k_0)$.

\item  $ \omega - {\mathbf U}_0 \cdot {\mathbf k}= \omega_D$ is the Doppler-shifted frequency in the moving frame, where the frequency of the signal  is reduced.  It is analogous to the reduction of frequency of the sound wave in a moving train when the train moves away from the source.  For $U_0=0$, it is customary to assume that $\omega \rightarrow 0$ since we focus on dynamics at large time scales~\citep{Yakhot:JSC1986,Zhou:PRA1988,McComb:book:Turbulence,Zhou:PR2010,McComb:book:new}.  The corresponding assumption is to set $\omega_D \rightarrow 0$ because $\omega_D$ is the effective frequency in the moving frame in which ${\bf U}_0=0$.  The approximation   $ \omega \rightarrow {\mathbf U}_0$ essentially takes away the effect of Galilean transformation and yields inherent turbulence properties.  Note that in Taylor's frozen-in turbulence hypothesis, $\omega = {\bf U_0 \cdot k}$ that yields $\omega_D = 0$ (see \S\ref{sec:Taylor}).  Therefore, 
\begin{eqnarray}
\delta \nu_{(0)}( k) k^2  =  \frac{1}{d-1}  \int_{\bf {p + q = k}}^{\Delta} \frac{d\mathbf p}{(2\pi)^d}
 \frac{B(k,p,q)  C(\mathbf p)}{   \nu_{(0)}(p) p^2 + \nu_{(0)}(q) q^2 },
 \label{eq:delta_nu3}
\end{eqnarray}
which is independent of $\mathbf U_0$.  The above formula is identical to that derived for ${\mathbf U}_0 = 0$.  Thus, in the Eulerian framework with Green's function and correlation function of the forms given by Eqs.~(\ref{eq:Green_fn_RG}, \ref{eq:correlation_fn_RG}),  the effects of the mean velocity field ${\mathbf U}_0$  disappears in the calculation.  This result however breaks down on inclusion of sweeping effect ($\tilde{\mathbf U}_0$) in Green's function; this is related to the sweeping effect of \citet{Kraichnan:PF1964Eulerian}. See the discussion in \S\ref{sec:sweeping_effect} and Eq.~(\ref{eq:G_k_omega_new}).

\begin{figure}
\begin{center}
\includegraphics[scale = 0.7]{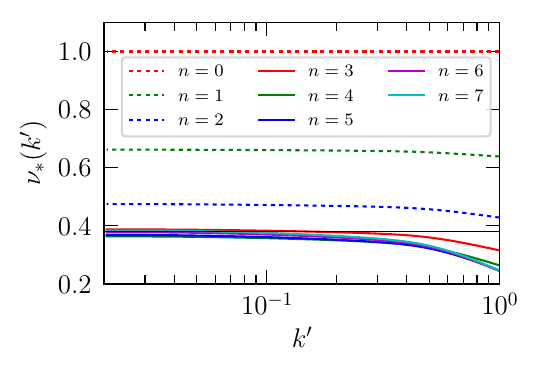}
\end{center}
\caption{Plot of  $\nu_*(k')$ vs $k'$.  The function approaches $\nu_*(k') \approx 0.38$ (the black horizontal line) asymptotically.}
\label{fig:RG}
\end{figure}

\item  The integral of Eq.~(\ref{eq:delta_nu3})  is performed over the first shell $(k_1,k_0)$.  Let us denote $\nu_{(1)}(k)$  as the renormalized viscosity after the first step of wavenumber elimination, i.e., 
\begin{eqnarray}
\nu_{(1)}(k) & = & \nu_{(0)}(k)+\delta\nu_{(0)}(k).  \label{eq:eta_1}
\end{eqnarray}
We keep eliminating the shells one after the other by the above procedure.   After $n+1$ iterations, we obtain 
\begin{eqnarray}
\nu_{(n+1)}(k) & = &  \nu_{(n)}(k)+\delta\nu_{(n)}(k)\label{nu_n},  \label{eq:nu_np1} \\
\delta \nu_{(n)}( k) k^2   & = &  \frac{1}{d-1}  \int_{\bf {p + q = k}}^{\Delta} \frac{d\mathbf p}{(2\pi)^d}
 \frac{B(k,p,q)  C(\mathbf p)}{   \nu_{(n)}(p) p^2 + \nu_{(n)}(q) q^2 },
 \label{eq:delta_nu_n}
\end{eqnarray}
with the integration performed over the $n$-th shell.

\item We compute Eqs.~(\ref{eq:nu_np1}, \ref{eq:delta_nu_n}) self-consistently.  We attempt Kolmogorov's energy spectrum for the energy, and obtain the renormalized viscosity iteratively (considering that the iteration procedure converges).   For the modal energy spectrum $C(\bf p)$, we substitute
\begin{equation}
C({\bf p}) =\frac{2(2\pi)^{d}}{S_{d}(d-1)}p^{-(d-1)}E(p);~~E(p) = K_\mathrm{Ko} \Pi^{2/3} p^{-5/3},
\end{equation}
where $S_{d}$ is the surface area of a $d$-dimensional sphere of unit radius, and $E(p)$ is the one-dimensional Kolmogorov's spectrum.  Regarding $\nu_{(n)}(k)$, we attempt the following form of solution
 \begin{equation}
\nu_{(n)}(k)  =   \nu_{(n)}(k_{n}k')   =  (K_\mathrm{Ko})^{1/2}\Pi^{1/3}k_{n}^{-4/3}\nu_{*(n)}(k')
\end{equation} 
 with  $k=k_{n}k'$ and $k'<1$.  The above equation is consistent with $\nu(k) \sim k^{-4/3}$.   We expect $\nu_{*(n)}(k')$ to be a universal functions for large $n$.  Substitutions of the above forms of $C({\bf p})$ and $\nu_{(n)}(k)$ in Eqs.~(\ref{eq:nu_np1}, \ref{eq:delta_nu_n})  yields the following equations: 

 \begin{eqnarray}
 \delta\nu_{*(n)}(k')  =    \frac{1}{(d-1)} && \int_{\bf{p'+q'=k'}}d\mathbf{q}'\frac{2}{(d-1)S_{d}}\frac{E^{u}(q')}{q'^{d-1}} \nonumber \\ && \times \left[ \frac{S(k',p',q')}{\nu_{*(n)}(hp')p'^{2}+\nu_{*(n)}(hq')q'^{2}}\right],
\label{eq:delta_nu*} \label{eq:nu'} 
\end{eqnarray}
\begin{eqnarray}
 \nu_{*(n+1)}(k')  =  h^{4/3}\nu_{*(n)}(hk')+h^{-4/3}\delta\nu_{*(n)}(k'), \label{eq:delta_nu'}
\end{eqnarray}

 where the integral in the above equation is performed  over a region $1\leq p',q'\leq1/h$ with the constraint  $\mathbf{p}'+\mathbf{q}'=\mathbf{k}'$. Note that $\mathbf{k}'= \mathbf{k}/k_n$, $\mathbf{p}'= \mathbf{p}/k_n$, $\mathbf{q}'= \mathbf{q}/k_n$. \citet{Fournier:PRA1978} showed the above volume integral in $d$ dimensions is 
 \begin{eqnarray}
 \int_{\mathbf{p}'+\mathbf{q}'  =  \mathbf{k}'}d\mathbf{p'} =S_{d-1} \int dp'dq'\left(\frac{p'q'}{k'}\right)^{d-2}\left(\sin\alpha\right)^{d-3},\label{eq:volume-integral}
\end{eqnarray}
where $\alpha$ is the angle between vectors $\mathbf{p}'$ and $\mathbf{q}'$.

\item $\nu_{*(n)}(k')$ is solved iteratively using Eqs.~(\ref{eq:nu'}-\ref{eq:delta_nu'}) with $h=0.7$~\citep{McComb:book:Turbulence,McComb:book:new,Zhou:PR2010}. \citet{Zhou:NASA1997} showed that the Kolmogorov constant computed using RG is approximately 1.6 independent of $h$, as long as it lies between 0.55 to 0.75.  Therefore we choose $h=0.7$ for our computation.  We start with a constant value of $\nu_{*(0)}(k')$, and compute the integral using Gaussian quadrature. This process is iterated till $\nu_{*(n+1)}(k')\approx \nu_{*(n)}(k')$, that is, till the solution converges. The result of our RG analysis, exhibited in Fig.~9, shows a constancy of $\nu_*(k')$ with $k'$.  A slight downward bend near $k'=1$ is attributed to the neglect of the triple nonlinearity of the unresolved modes (see item 1,  and \citet{Zhou:PRA1988}).
\end{enumerate}

For large $n$,   $\nu_{*(n)}(k')$ converges asymptotically to $\nu_* \approx 0.38$ as $k'\rightarrow 0$.  The above result is same as that for ${\mathbf U}_0=0$, thus we conclude that the renormalized viscosity $\nu_n(k)$ is independent of $\mathbf U_0$.

The aforementioned description does not fully match with the numerical simulation described in \S\ref{sec:sweeping_effect} and \S\ref{sec:Taylor}.  In the numerical simulations we showed that the absolute value of the normalised correlation function or the Green's function decays in time with time scale $1/\nu(k) k^2$.  However the imaginary part and phase of the correlation function exhibit fluctuations that are related to the sweeping by eddies of larger scales, as predicted by \citet{Kraichnan:PF1964Eulerian}.  This sweeping effect is not captured by the Eulerian field theory described above, which was first argued by \citet{Kraichnan:PF1964Eulerian}.  \citet{Kraichnan:PF1965Lagrangian_history} then formulated Lagrangian-history closure approximation for turbulence  and showed consistency \citep[also see][]{Leslie:book}.  Also refer to \citet{Moriconi:JSM2014} and \citet{Okane:PS2008} for more work on field-theoretic treatment of turbulence.  We do not describe this formalism due to limited scope of the present paper.  


\end{document}